\documentstyle[12pt,axodraw]{article}
\begin{document}
\hfill{NCKU-HEP/95-05}
\hfill{October 7 1995}
\begin{center}
\vspace{5mm}
{\large \bf Structural Aspects of the Proton in the Chiral Quark Model\\
}
\vspace{15mm}
Chung-Yi Wu\\
\vspace{5mm}
Department of Physics\\
National Cheng Kung University\\
Tainan, Taiwan 701, Republic of China\\
\end{center}
\begin{center}
{\bf ABSTRACT\\}
\end{center}

We calculate with chiral symmetry the parton contents of the
proton based on a two-component wave function.  The
calculation results give significant sea-quark contents
and, especially, the intrinsic gluon polarization produced
at a more fundamental level.
\\
\begin{center}
\end{center}

\newpage
How the spin of the proton is distributed among its
constituents, quarks and gluons, is an important problem in
hadron physics.  An enormous experimental effort has been
undertaken to provide precise information on this problem.  In
particular, measurements of polarization correlations in
high momentum transfer reactions provide highly sensitive tests
of the underlying structure and dynamics of hadrons.

For the past seven years, there have been
several theoretical investigations on the substructure of the nucleon
inspired by the EMC measurement [1] of the polarized
proton structure function $g_1^p$.  The inspiration in spin physics
mainly originated from the EMC results, which imply that within the
na{\"\i}ve parton model the sum of spins carried by all quarks
and antiquarks almost vanishes, in
contradiction to the expectation of the model [2] with
sum of spins being $\frac{1}{2}$.  In this letter, we shall
show that the discrepancy can be resolved
in the chiral quark model [3] with a two-component proton
wave function [4] by giving an account of the flavor and spin
contents of the proton.

Since most processes involve both low and high energy
aspects, one should separate the low and high energy pieces in a
multiplicative way via the idea of factorization.  While
in the high energy regime the Altarelli-Parisi
equations [5] govern the $Q^2$-evolution
($Q$: the momentum transfer) of the structure
functions, non-perturbative techniques must still
be developed to attempt any kind of description of the elementary
features on the flavor and spin contents of the
nucleon.  An important step based on the effective chiral quark
theory is provided by Manohar and Georgi [3].  The successes
of the chiral symmetric description of low energy
hadron physics indicate that the chiral symmetry scale
$\Lambda_{\chi SB} \sim 1GeV$ is higher
than the QCD confinement scale, and thus the important
degrees of freedom at momentum scales relevant to hadron structure
should be quarks, gluons and Nambu-Goldstone bosons.  In
the chiral quark model, the dominant process for the flavor
content corrections is the dissociation of a quark into a quark
and a Nambu-Goldstone boson, Fig.1; internal gluon effects are
negligibly small [3, 6-9].
\begin{center} \begin{picture}(300,56)(0,0)
\ArrowLine(50,10)(125,10)
\Text(80,3)[]{$q$}
\ArrowLine(125,10)(200,10)
\Text(170,3)[]{$q'$}
\DashLine(125,10)(140,30){3}
\ArrowLine(140,30)(170,20)
\Text(180,20)[]{$\overline{q'}$}
\ArrowLine(140,30)(170,40)
\Text(180,40)[]{$q$}
\end{picture}
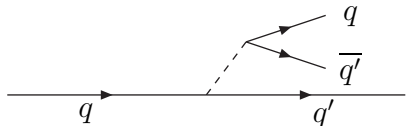
\\{Fig.1: Fluctuation of a quark in chiral field theory.}
\end{center}

The essential step of our investigation is to consider the
two-component proton [4]
\begin{equation}
\mid\rm p\rangle=\mit \cos\theta\mid 3q\rangle
+\sin\theta\mid (3q)G\rangle\,,
\end{equation}
where in the second component a $3q$ color-octet wave function with
spin and isospin $\frac{1}{2}$ is coupled with a spin-1 color-octet
gluon $G$ to make a color-single state with total angular momentum
$J=\frac{1}{2}$.  The angle $\theta$ specifies the amount
of mixing.  In this physical picture, the gluon density can be
regarded as an intrinsic quantity.

We now calculate the proton's
flavor contents with a broken-U(3) symmetry [9].  The U(3) symmetry
is broken at the subleading contributions in the $1/N_c$
expansion (non-planar diagrams) [10], where $N_c$ is the number
of colors.  The broken-U(3) can be implemented by
taking the differently-renormalized octet
and singlet Yukawa couplings, $\mit g_0/g_8\equiv \zeta \neq$ 1
[9].  The interaction vertex is
\begin{equation}
\cal L_{\mit I}= \mit g_8 \overline{q} \it\widetilde{\phi}\mit q + \sqrt
\frac{2}{3} g_0 \overline{q}\eta'q\,,
\end{equation}
where
\begin{eqnarray*}
\widetilde{\phi}=\sum_{i=1}^{8} \lambda_i \phi_i= \left(
 \begin{array}{ccc}
  \frac{\pi^0}{\sqrt 2}+\frac{\eta}{\sqrt 6} & \pi^+ & K^+ \\
  \pi^- & \frac{-\pi^0}{\sqrt 2}+\frac{\eta}{\sqrt 6} & K^0 \\
  K^- & \overline{K}^0 & \frac{-2\eta}{\sqrt 6} \\
 \end{array}
 \right)\,,
\end{eqnarray*}
$q=(u,d,s)$ and $\lambda^,s$ are the Gell-Mann matrices.  The
ninth singlet $\eta'$ and $\eta$ are unphysical.  In this model
we can express the flavor densities in terms of three
parameters $\theta$, $\zeta$ and $a$, where $a$ is the
probability of a $\pi^+$ emission.  We suppose that the
Goldstone boson fluctuation is sufficiently small to be
treated as a perturbation.  The probabilities
of no Goldstone emission are $\cos^2\theta[1-\frac{a}{3}
(\zeta^2+8)]$ and $-\frac{1}{3}\sin^2\theta[1-\frac{a}{3}
(\zeta^2+8)]$ corresponding to the first and second component
in (1), respectively.  The multiplicative factor $-\frac{1}{3}$
appearing in the second probability comes from the fact that
the total quark spin in the hybrid $3qG$ state
has a probability of $\frac{1}{3}$ of being parallel to the
total spin.  After one interaction, the
proton's antiquark contents read
\begin{eqnarray}
\overline{u}&=&(\cos^2\theta-\frac{1}{3}\sin^2\theta)[\frac{a}{3}(
\zeta^2+2\zeta+6)]\,,\\
\overline{d}&=&(\cos^2\theta-\frac{1}{3}\sin^2\theta)[\frac{a}{3}(
\zeta^2+8)]\,,\\
\overline{s}&=&(\cos^2\theta-\frac{1}{3}\sin^2\theta)[\frac{a}{3}(
\zeta^2-2\zeta+10)]\,,
\end{eqnarray}
and the quark contents are $u=2+\overline{u}$, $d=1+\overline{d}$ and
$s=\overline{s}$.

We now turn to the proton's spin contents.  A quark can change its
helicity by emitting a spin zero meson, Fig.2.
\begin{center} \begin{picture}(300,56)(0,0)
\ArrowLine(0,20)(40,20)
\Text(10,13)[]{$u_\uparrow$}
\ArrowLine(40,20)(80,20)
\Text(70,13)[]{$d_\downarrow$}
\DashLine(40,20)(55,40){3}
\Text(65,45)[]{$\pi^+$}
\Text(40,7)[]{(a)}
\ArrowLine(100,20)(140,20)
\Text(100,13)[]{$u_\uparrow$}
\ArrowLine(140,20)(180,20)
\Text(170,13)[]{$s_\downarrow$}
\DashLine(140,20)(155,40){3}
\Text(165,45)[]{$K^+$}
\Text(140,7)[]{(b)}
\ArrowLine(200,20)(240,20)
\Text(210,13)[]{$u_\uparrow$}
\ArrowLine(240,20)(280,20)
\Text(270,13)[]{$u_\downarrow$}
\DashLine(240,20)(255,40){3}
\Text(265,45)[]{$\pi^0,\eta,\eta'$}
\Text(240,7)[]{(c)}
\end{picture}
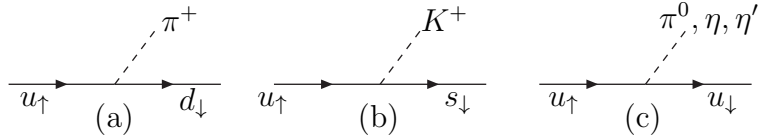
\\{Fig.2: A valence up quark changes its helicity by emitting a
spin zero meson.}
\end{center}
After one interaction, the contributions of various spin states
can be read off from the proton's flavor composition
\begin{eqnarray*}
&&\cos^2\theta \left\{[1-\frac{a}{3}(\zeta^2+8)](\frac{5}{3}u_\uparrow+
\frac{1}{3}u_\downarrow+
\frac{1}{3}d_\uparrow+\frac{2}{3}d_\downarrow)\right.\\
&&\left.+\frac{5}{3}\mid\Psi(u_\uparrow)\mid^2+
\frac{1}{3}\mid\Psi(u_\downarrow)\mid^2+\frac{1}{3}\mid\Psi
(d_\uparrow)\mid^2+ \frac{2}{3}\mid\Psi(d_\downarrow)\mid^2 \right\}\\
&&-\frac{1}{3}\sin^2\theta \left\{ [1-\frac{a}{3}(\zeta^2+8)]
(\frac{4}{3}u_\uparrow+\frac{2}{3}u_\downarrow+\frac{2}{3}d_\uparrow
+\frac{1}{3}d_\downarrow)\right.\\
&&\left.+\frac{4}{3}\mid\Psi(u_\uparrow)\mid^2+\frac{2}{3}\mid\Psi
(u_\downarrow)\mid^2+ \frac{2}{3}\mid\Psi(d_\uparrow)\mid^2
+\frac{1}{3}\mid\Psi(d_\downarrow)\mid^2 \right\}\,,
\end{eqnarray*}
where
\begin{eqnarray*}
\mid\Psi(u_\uparrow)\mid^2&=&(\frac{2}{3}a+\frac{1}{3}a\zeta^2)
u_\downarrow +ad_\downarrow+as_\downarrow\,,\\
\mid\Psi(d_\uparrow)\mid^2&=&(\frac{2}{3}a+\frac{1}{3}a\zeta^2)
d_\downarrow +au_\downarrow+as_\downarrow\,,
\end{eqnarray*}
with $\mid\Psi(u_\downarrow)\mid^2$ and $\mid\Psi(d_\downarrow)\mid^2$
having the same but opposite helicity expressions.  Thus the quark
contributions to the proton spin $\Delta q$ = $q_\uparrow
-q_\downarrow+ \overline q_\uparrow -\overline q_\downarrow$ are :
\begin{eqnarray}
\Delta u&=& \cos^2\theta[\frac{4}{3}-\frac{1}{9}(8\zeta^2+37)a]\nonumber\\
&&+\sin^2\theta[-\frac{2}{9}+\frac{1}{27}(4\zeta^2+23)a]\,,\\
\Delta d&=& \cos^2\theta[-\frac{1}{3}+\frac{2}{9}(\zeta^2-1)a]\nonumber\\
&&+\sin^2\theta[-\frac{1}{9}+\frac{2}{27}(\zeta^2+8)a]\,,\\
\Delta s&=& -(\cos^2\theta-\frac{1}{3}\sin^2\theta)a\,,
\end{eqnarray}
which lead to the total quark spin contribution
\begin{eqnarray}
\Delta\Sigma&=&\Delta u+\Delta d+\Delta s \nonumber\\
&=&(\cos^2\theta-\frac{1}{3}\sin^2\theta)[1-\frac{2}{3}(\zeta^2+8)a]\nonumber\\
&=& \cos^2\theta-\frac{1}{3}\sin^2\theta -2\overline d\,,
\end{eqnarray}
and the intrinsic gluon polarization
\begin{eqnarray}
\Delta G = \frac{2}{3}\sin^2\theta+\overline d\,,
\end{eqnarray}
where $\Delta G=G^\uparrow-G^\downarrow$.  The
general spin decomposition of the proton is
$\frac{1}{2}=\frac{1}{2}\Delta\Sigma+\Delta G+L_z$, with  the orbital
angular momentum $L_z$ being taken to vanish in the ground
state.  In terms of the three parameters $\theta$, $\zeta$, and
$a$, we obtain the general expressions (3)-(10)
for flavor and spin densities.

Our model calculations can reproduce the results of
Refs. [4] , [8] and [9] by adjusting the parameters.  When
$\sin^2\theta=0$, we can obtain
the results of the SU(3) octet ($\zeta=0$), the results of the U(3) symmetry
theory ($\zeta=1$) [8] and the expressions in Ref. [9]
($\zeta \neq 1$); the non-vanishing gluon polarization (10)
corresponds to the residual proton spin excluding the total
quark spin contribution in Refs. [8] and [9].  Taking $a=0$ (no
Goldstone boson emission) gives the
expressions of Lipkin [4].

It is also instructive to look back
at the SU(3) octet and the U(3) symmetry
theory with $\sin^2\theta\neq0$.  From (3) and (4), one has
$\overline{u}/\overline{d}=0.75$ for the SU(3) octet and
$\overline{u}/\overline{d}=1$ for the U(3) symmetry case, while
NA51 asymmetry measurement [11] gives
\begin{eqnarray}
\overline{u}/\overline{d}=0.51\pm0.04(stat)\pm0.05
(sys)\,.
\end{eqnarray}
Clearly, the two-component proton wave function gives the
same values $\overline{u}/\overline{d}=0.75$ and
$\overline{u}/\overline{d}=1$, which are the same as those obtained
from the original SU(6) proton wave function, do not agree with
the NA51 experimental observation.  The
main hint of the $\overline{u}-\overline{d}$ asymmetry comes
from the violation of the Gottfried sum rule [12], which
is provided by the NMC measurement [13] of $S_G=0.235\pm0.026
(Q^2=4GeV^2)$.  Under
the assumptions of isospin symmetry for the nucleon and for the sea
quark distributions in the proton, the sum rule reads
\begin{eqnarray*}
 S_G &=& \int_0^1{dx\over x}\left[F_2^{\mu p}(x)
                            -F_2^{\mu n}(x)\right]\nonumber\\
     &=& {1\over 3}\int_0^1dx\left[u(x)+\overline{u}(x)-d(x)
                            -\overline{d}(x)\right]\nonumber\\
     &=& {1\over 3}\,.
\end{eqnarray*}
It implies a large SU(2) flavor symmetry breaking in the quark
sea or a suppression for the production of $u{\bar u}$ pairs
relative to $d{\bar d}$ pairs in the proton,
\begin{eqnarray}
 \overline{d} -\overline{u} &=&
\frac{2}{3}(\cos^2\theta - \frac{1}{3}\sin^2\theta)(1-\zeta)a \nonumber\\
&=&0.1475\mp0.039\,.
\end{eqnarray}
As a rough estimate, by combining (11) with (12) and taking the
current value $\Delta \Sigma=0.27$ [14] for (9), we get
$\sin^2\theta=0.1$.  This
value is different from that of Ref. [4], where
$\sin^2 \theta=\frac{15}{64}$ is fitted from the Bjorken
sum rule [15].  The difference arises from a
modification of the parton contents by including the Goldstone boson
fluctuation.

On the other hand, the experimental uncertainties from the
statistical and systematic errors give a large freedom of
choices on these parameters. We now
illustrate our model calculations with the following
simple choice of parameters:
\begin{equation}
\sin^2 \theta=0.05\,,~a=0.09 \,,~\zeta=-1\,,
\end{equation}
where we see that the contribution from the second component of
the proton wave function is suppressed.  The parton contents (3)-(10)
yield the values :
\begin{eqnarray}
\Delta u&=&0.83\,,~
\Delta d=-0.32\,,\nonumber\\
\Delta s&=&-0.08\,,~
\Delta G=0.29\,,\nonumber\\
\overline u/\overline d&=&0.556\,,~
\overline{d}-\overline{u}=0.112\,,
\end{eqnarray}
and the fractions of quarks flavor $f_a\equiv(q_a+\overline{q_a})/
\Sigma(q+\overline q)$ read
\begin{eqnarray}
f_u=0.51\,,~f_d=0.33\,,~f_s=0.16\,.
\end{eqnarray}
Our results give significant strange-quark content.  With
the phenomenological value $\sigma_{\pi N}\approx 45 MeV$ [16]
extracted from the isospin symmetric part of the $\pi N$
scattering amplitude, Cheng and Li [9] deduced
$(f_s)_{\sigma_{\pi N}}=0.18$.  Within these model dependent
analyses, the matrix element of the strange scalar operator in
the nucleon $\langle N\mid\overline ss\mid N\rangle$ is
not negligible.  In addition, the magnetic
moments in quark magnetons are given by
\begin{eqnarray*}
\mu_p=0.687\,,~\mu_n=-0.463\,,
\end{eqnarray*}
with a ratio
\begin{equation}
\mu_p/\mu_n=-1.484\,,
\end{equation}
in the flavor-SU(3) limit of $m_{u,d}=m_s$.  This
is closer to the experimental value $-1.46$ than the original
SU(6) prediction $\mu_p/\mu_n=-\frac{3}{2}$.

We have shown that a simple extension in the chiral quark
model enjoys many interesting features.  For example,
the intrinsic gluon polarization can be drawn at a more
fundamental level.  In general, the sea-quark or anomalous
gluonic interpretation for the violation of the Ellis-Jaffe sum
rule depends on the factorization scheme defined
for the quark spin density and the cross section for the photon-gluon
scattering.  In perturbative QCD, the polarized
proton structure function is expressed as
\begin{eqnarray}
g_1^p(x,Q^2) &=& {1\over 2}\sum_i^{n_f} e^2_i \left\{\int_x^1 \frac{dy}{y}
\Delta q_i(y,Q^2)\times [\delta (1-\frac{x}{y}) + \frac{\alpha_s(Q^2)}{2\pi}
\Delta f_q(\frac{x}{y})]\right.\nonumber \\
&&\left.-\frac{\alpha_s(Q^2)}{2\pi}\int_x^1\frac{dy}{y}
\Delta \sigma^{hard}(\frac{x}{y})\Delta G(y,Q^2)\right\}\,,
\end{eqnarray}
where
\begin{eqnarray*}
\Delta f_q(z)&=&f_q(z)-\frac{4}{3}(1+z)\,,\nonumber \\
\int_0^1f_q(z)dz&=&0\,,~
\int_0^1\Delta f_q(z)=-2\,,
\end{eqnarray*}
and the hard kernels [17]
\begin{eqnarray}
\Delta \sigma^{hard} (x)&=&(1-2x)(\ln\frac{Q^2}{\mu^2_{fact}}
+\ln\frac{1-x}{x}-1)-2(1-x)\,,\nonumber \\
\Delta \widetilde{\sigma}^{hard} (x)&=&(1-2x)(\ln\frac{Q^2}{\mu^2_{fact}}
+\ln\frac{1-x}{x}-1)\,,
\end{eqnarray}
with
\begin{eqnarray*}
\int_0^1 \Delta\sigma^{hard} (x)=0\,,~
\int_0^1 \Delta\widetilde{\sigma}^{hard} (x)=1\,,
\end{eqnarray*}
where $\Delta \sigma^{hard} (x)$ and $\Delta \widetilde{\sigma}^{hard}
(x)$ correspond to gauge-invariant factorization scheme and
chiral-invariant (gauge-variant) scheme [18],
respectively.  In perturbative QCD, the gluonic contribution
depends on the hard kernels (18); a different choice
of kernels will yield a different set of parton distributions.  We
assume $\Delta u_s=\Delta d_s=
\Delta s$ and $\Delta q_v=\Delta q'_v$ [19], where
$\Delta q$ and $\Delta q'$ correspond to
gauge-invariant and chiral-invariant parton
spin densities, respectively.  Then from (17) and (18)
with their first moments, one gets
\begin{equation}
\Delta s = \Delta s' - \frac{\alpha_s}{2\pi}\Delta G\,.
\end{equation}
We now take the threshold momentum transfer $Q=1GeV$
($\alpha_s=0.434$, $\Lambda_{QCD}=200MeV$) to relate (19)
with the model calculation results.  It is
straightforward to see that there is a good agreement with
relation (19) with $\Delta s=\Delta s_{exp}
=-0.10$, $\Delta s'=-0.08$ and $\Delta G=0.29$ taken from
(14).  We will get poor agreement if the initial
momentum transfer is below the threshold
value, since perturbative QCD is not applicable
in region below the threshold momentum transfer.  Above
the threshold scale, the chiral symmetry breaking
occurs and the model dependent results
are not applicable.

With regard to the proton spin
crisis, suggestions have been made in terms of a large
negative polarization of the sea quarks inside the proton [6, 20, 21]
or a large positive polarization for the gluons [22].  A
suitable combination of both should be
more realistic. Our model calculations give a sensible
combination on the first moments of the parton
distributions.  Further analyses would require the $Q^2$- and
$x$-dependent parton distributions in the
perturbative region.

In using the Altarelli-Parisi equations [5], one has to know the
initial behaviors of $Q^2$ and $x$ dependences
of the gauge-invariant parton distributions.  In particular, the
polarized parton distributions are based on the
parameterizations with theoretical
prejudice.  To date, the only two constraints on the
distributions are their first moments and the requirements of
positivity of the spin-parallel and spin-antiparallel
distributions [23].

Thus it is worthwhile
to embark on a more detailed theoretical
investigation, especially, the
polarized parton distribution functions.  The
presented nucleon substructure
from the chiral quark model may serve as a basis for
further studies.

The author is indebted to Prof. Su-Long Nyeo for suggestions
and encouragement.  He wishes to thanks  Prof. Yeou-Wei Yang for
several discussions.  This research was supported by the National
Science Council of the Republic of China under Contract No.~NSC
85-2112-M006-004.

\end{document}